# Crystal channeling of LHC forward protons with preserved distribution in phase space

V.M. Biryukov[♦]

*Institute for High Energy Physics, Protvino, 142281, Russia*

**Abstract**

We show that crystal can trap a broad (*x, x', y, y', E*) distribution of particles and channel it preserved with a high precision. This sampled-and-hold distribution can be steered by a bent crystal for analysis downstream. In simulations for the 7 TeV Large Hadron Collider, a crystal adapted to the accelerator lattice traps 90% of diffractively scattered protons emerging from the interaction point with a divergence 100 times the critical angle. We set the criterion for crystal adaptation improving efficiency ~100-fold. Proton angles are preserved in crystal transmission with accuracy down to 0.1 μrad. This makes feasible a crystal application for measuring very forward protons at the LHC.

**1. Introduction**

A particle entering the crystal lattice parallel to a major crystallographic direction can be captured and channeled by the lattice along a crystal axis or plane [1]. For instance, a positive particle can be channeled between adjacent atomic planes. In a bent crystal, the channeled particles can follow the bend [2]. This led to elegant technique of beam steering by bent channeling crystals [3] now experimentally explored over six decades in energy from low MeV [4] to 1 TeV [5]. The technique is used on permanent basis in IHEP Protvino where crystal systems extract protons from 70 GeV main ring with efficiency of 85% at intensity up to $4\times10^{12}$ protons using Si crystals just 2 mm along the beam [6]. Bent crystals channel in good agreement with predictions up to the highest energies [6-9].

Crystal applications at the 7-TeV Large Hadron Collider are considered for beam collimation and extraction [10] and *in situ* calibration of CMS and ATLAS calorimeters [11]. In another proposal, crystal could capture the particles emerging from the interaction point (IP) with small angles and channel them out of the beam [12]. This could help to improve on measurement of small angle elastic and "quasi-elastic" scattering in CMS and ATLAS where lower momentum transfers might become available for *pp* elastics scattering and lower proton momentum losses for diffractive physics [12]. Groups in both CMS (with TOTEM) and ATLAS would like to add very forward proton detectors, 420 m downstream on both sides, a project FP420 [13]. By detecting protons that have lost less than 1% of their longitudinal momentum, a rich QCD, electroweak, Higgs and BSM program becomes accessible, with the potential to make measurements which are unique at LHC, and difficult even at a future linear collider [13].

The measurement of the displacement $x$ and angle $x'$ (in the horizontal plane) of the outgoing protons relative to the beam allows the momentum loss $\xi=\Delta p/p$ and transverse momentum of the scattered protons to be reconstructed. Protons emerging from diffractive scattering at LHC have very small emission angles (10-150 µrad) and fractional momentum loss ($\xi = 10^{-8} - 0.1$). Hence they are very close to the beam and can only be detected in the Roman Pots downstream if their displacement at the detector location is large enough to escape the beam halo [13].

## 2. Crystal efficiency

As practice shows, crystal can go into a very limited space and get particles from there [6]. Most efficient crystal applications are based on so called "multipass" mode where particles can encounter a crystal many times in the ring [6,14]. There are also successful experimental demonstrations of highly efficient channeling in a single pass, with efficiency up to 60% at CERN SPS [15]. Throughout this paper we consider only a single-pass channeling.

We show with simulations that at the LHC a crystal can efficiently channel forward protons. For channeling simulations we apply a Monte Carlo code CATCH [16] successfully used for prediction of experiments at CERN SPS [9], IHEP U-70 [6], Tevatron [7], RHIC [8] and KEK [17] and crystal applications at the LHC [10,11].

Crystal capture is very selective in angle. The critical angle $\theta_C$ within which the capture is possible is as small as about ±5 µrad / $E^{1/2}$(TeV) at a high energy $E$ in Silicon (110) planes. Proton divergence of 150 µrad is almost 100 times $\theta_C$ at 7 TeV. Therefore it is not possible to capture all these protons by a plain crystal. However, we can suggest an efficient solution benefiting from the fact that all diffractively scattered protons originate from a small region at the IP.

For standard LHC optics with beta function value at the IP $\beta^*$=0.55 m, the beam size at the IP is $\sigma_{beam}$≈16 µm rms. The spread in the transverse position of the vertex point where outgoing protons originate from is determined by the rms spread of the beam and equals $\sigma_{beam}/\sqrt{2}$ [13]. As protons emerge from diffractive scattering at LHC with emission angles up to ≈150 µrad and interaction width $\sigma_{beam}/\sqrt{2}$, the emittance of the beam to be trapped by a crystal is ≤ 2π µrad-mm only. This corresponds to the acceptance of a Si crystal of ~1 mm transverse size. The match of the diffractive-protons emittance to the crystal acceptance means that the particles could be

---
♦ http://mail.ihep.ru/~biryukov/

trapped and channeled efficiently. To realize this, one has to match the crystal design and location to the application. For the LHC scenario with high luminosity, we find the most efficient design to be a crystal with a *point-to-parallel focusing* entry face. Focusing crystal proposed by A.I Smirnov in 1985 has a face shaped so that the tangents to the crystal planes cross at a focus line at some distance $L_F$ from the crystal [3]. This kind of crystal was successfully tested at IHEP where it efficiently trapped a beam of ±2 mrad divergence (or ~100 times $\theta_C$) [18]. The crystal traps protons emerging from the focus line uniformly from all the angular range if the entry face has a proper shape [18].

We find that for most efficient channeling in the LHC the focus distance $L_F$ of this crystal should be equal to the effective distance $L_{eff}$ between the crystal location and the IP. In a drift space, $L_{eff}$ is geometrical distance. In accelerator lattice, $L_{eff} = (\beta^*\beta_C)^{1/2}\sin\Delta\psi$, where $\beta_C$ is β value at the crystal and $\Delta\psi$ is the phase advance from the IP to the crystal. A plain crystal with a flat entry face has $L_F = \infty$.

In simulations with the low $\beta^*$ optics settings, a focusing crystal shows best efficiency if installed at a location with effective distance $L_{eff} \geq 15$ m from the IP. It can be a Si (110) or (111) crystal of ≈(0.15 mrad)×$L_{eff} \geq 2.5$ mm transverse size in order to capture efficiently all diffractive protons. Simulation predicts that a focusing Si(110) crystal with $L_F=L_{eff}$ traps 90% of 7 TeV protons emerging from the IP in the angular range of 150 μrad width into channeling mode. The efficiency figure is almost independent of the crystal location provided $L_{eff} \geq 15$ m.

The reason for high efficiency at high $L_{eff}$ is that, at a distance $L_{eff}$ from the IP, any point at the crystal entry face sees the beam source of σ size (at the IP) at an angle of σ/$L_{eff}$. Channeling efficiency reduces by a factor of about $(1-(\sigma/L_{eff}\theta_C)^2)^{1/2}$ [3]. The reduction in efficiency by a factor of ≈$1-\sigma^2/2L^2_{eff}\theta_C^2$ becomes negligible for $L^2_{eff} >> \sigma^2/2\theta_C^2 = \beta^*\epsilon/4\theta_C^2$ where ε is beam emittance. With $\beta^*$=0.55 m and Si(110) crystal, channeling efficiency saturates for $L^2_{eff} >> (4$ m$)^2$.

One more idea for efficient channeling of forward protons in the LHC is that a crystal can be installed with planes parallel to either *x'* or *y'* plane. For application, it is not critical whether crystal bends protons in horizontal or vertical plane to produce an offset at the detector. But the distance $L_{eff}$ in accelerator lattice from the IP to the crystal can be very different in *x* and *y* planes. Then channeling efficiency is very different, whether crystal traps protons in *x'* or *y'* plane. We suggest in this case to install crystal for channeling in the plane with larger $L_{eff}$.

For instance, on the location 200 m downstream of the IP5 (CMS) and some 20 m ahead of the Roman Pot station at 220 m where crystal could be installed, $L_{eff}$ ≈6 m in *x'* and ≈20 m in *y'* plane. According to the analysis above, channeling efficiency in *y'* plane should be great while in *x'* plane moderate. Indeed, our simulations for this location show channeling efficiency of ≈87% in $y'$ and only ≈60% in $x'$ plane, for $\beta^*=0.55$ m and optimal crystals of Si(110). A plain Si crystal has channeling efficiency in $x'$ plane of just 3.5% or 17 times lower than a Si focusing crystal adapted to the LHC lattice.

For the run-in phase of the LHC with $\beta^*=2$ m we find that channeling efficiency of ≥ 85% can be achieved if crystal is located at $L_{eff} \geq 30$ m downstream of the IP. The nominal, high luminosity optics of the LHC is not optimized for forward proton detection. Therefore a possibility to use a channeling crystal can be very helpful as it offers opportunities for diffractive physics studies otherwise inaccessible in the nominal LHC settings.

The LHC options with a high $\beta^*$ (1540 and 90 m) are devised for the studies of diffractive physics. With $\beta^*=1540$ m, the emittance of diffractively scattered protons increases to ≈50π μrad-mm. This corresponds to the acceptance of a Si crystal of ≥ 30 mm transverse size. Such a crystal is not out of question, however the problem is where to fit it in the LHC. In terms of $L_{eff}$, good channeling efficiency requires a location with $L^2_{eff} >> (130\ m)^2$ in this optics.

We simulated channeling on the location 200 m from the IP5. In $\beta^*=1540$ m optics, a 10-cm Si(110) crystal trapped and bent protons 0.5 mrad in $x'$ plane with efficiency of 41%. A Ge(110) crystal shows there 48% efficiency, i.e. comparable to Si. All figures assumed a perfect match $L_F=L_{eff}$ in crystal. A plain Si crystal gives efficiency of <<1%, or 300 times lower than a Si focusing crystal adapted to the LHC lattice on this location.

In $\beta^*=90$ m option on the same location, the choice of plane is important because $L_{eff} \approx 10$ m in $x'$ and ≈170 m in $y'$ plane. Preferred location should have $L^2_{eff} >> (60\ m)^2$ so we expect very different efficiencies in $x'$ and $y'$ planes. Our simulations give crystal efficiency of 72% in $y'$ and only 7% in $x'$ plane for $\beta^*=90$ m. Here, crystal application is feasible only with bending in vertical plane.

Low efficiency may exclude a crystal use for double-pomeron-exchange events ($pp \rightarrow pXp$) with double-arm reconstruction, because the probability to have channeling in both arms in coincidence becomes small, e.g. $(41\%)^2 \approx 17\%$. For reconstruction of single-diffraction events ($pp \rightarrow pX$) more detailed studies are required before the benefits (or their absence) from a crystal use with high $\beta^*$ options can be understood.

In this paper we suggest the use of a *single* crystal for proton extraction from halo and delivery to the detector. The use of a *2-stage* crystal system [12], first crystal for extracting a proton and second one for bending it a big angle, would reduce the overall efficiency by a factor of ~0.6 (ideally) or less. The 2nd crystal traps only part of the protons channeled in the 1st one.

Finally, we notice that one can filter diffraction events with a crystal. Instead of trapping all forward protons, crystal acceptance can be made smaller and sample e.g. only the most forward protons emerging from the IP with the angles of a few μrad.

## 3. Precise transmission in a single (x, x') plane

Whereas protons are physically delivered from the IP to detector with good efficiency, the essential question is whether the information on phase space (x, x', y, y', E) distribution of particles is lost or corrupted while the particles are captured and transmitted in crystal. The success of experiments on measuring forward high momentum protons at the LHC depends on the angular precision of proton track reconstruction. A plain crystal would destroy the phase space information first by selecting particles from just a single direction and then disturbing the exit angle of particle by coherent and incoherent scattering in crystal. Plain crystal acceptance is $\pm\theta_C$ and crystal accuracy in angle transmission is again $\pm\theta_C$. That means, a plain crystal traps and delivers about zero bit information on angle distribution. In this paper we design a crystal with the acceptance of ~100 $\theta_C$ and angle transmission accuracy of ~0.1 $\theta_C$, although it sounds against the nature of crystal channeling.

Suppose particles are coming with a distribution over (x, x', y, y', E). Ideally, we would like the crystal to trap all coming particles and preserve their distribution over (x, x', y, y', E), and then shift an angle of $\theta$ each particle towards a physical setup where this distribution can be analyzed in detail.

One should solve two problems. One problem was to trap and bend a beam with a divergence much greater than the critical angle. A focusing crystal adapted to the LHC optics solves this problem. In simulations, a focusing crystal traps with 90% efficiency all protons emerging from the IP with the angular distribution ~100 times $\theta_C$. Notice that the trapped particles fully preserve also their distribution over the angle in the plane orthogonal to the plane of channeling. In such a crystal, particles are trapped uniformly from a very broad distribution over x' and y'.

A bent crystal would transform (x, x', y, y') at the entrance into (x, x'+$\theta$, y, y') at the exit. To do so, each trapped particle has to be channeled over the same distance in crystal. Therefore, the shape of the crystal exit face must match the entry face. Then in a bent crystal each channeled particle receives the same bending angle.

Although the crystal described above can solve the idea of sampling a broad distribution of particles and delivering it to a required destination, second problem is how to preserve the sampled distribution (x, x', y, y') "frozen" on transmission through the crystal lattice as precise as possible. The coordinates (x, y) of particles are obviously preserved in crystal, so one should take care of the accuracy in transmission of angles x', y' only.

The protons channeled between atomic planes in crystal are disturbed by (1) oscillations in the channeling plane with an amplitude up to $\theta_C$ and by (2) scattering on a rarefied electronic gas (mostly valence electrons) in both planes, x' and y'. Notice that nuclear scattering will not disturb the sample of transmitted channeled particles as this process is strongly suppressed for channeled positive particles. Simply saying, any particle nuclear scattered would be dechanneled and thus not present in the sample of bent particles.

That gives us the first idea that partially solves the problem of transmission accuracy. The idea is that the information on crystal-captured particles is very well preserved in one plane, e.g. $(x, x')$, while the particles are trapped and bent in another plane, e.g. $(y, y')$. Notice that particle distribution in the plane orthogonal to channeling is favored twice. Firstly, they are easily trapped with a broad angular distribution; secondly they are transmitted with a very little scattering. Information in this plane will be best preserved. The opportunity to have perfect data on just one plane is interesting for applications. The reconstruction of the Higgs boson mass in reaction $pp \rightarrow p+H+p$ requires $(x, x')$ data in horizontal plane only [13].

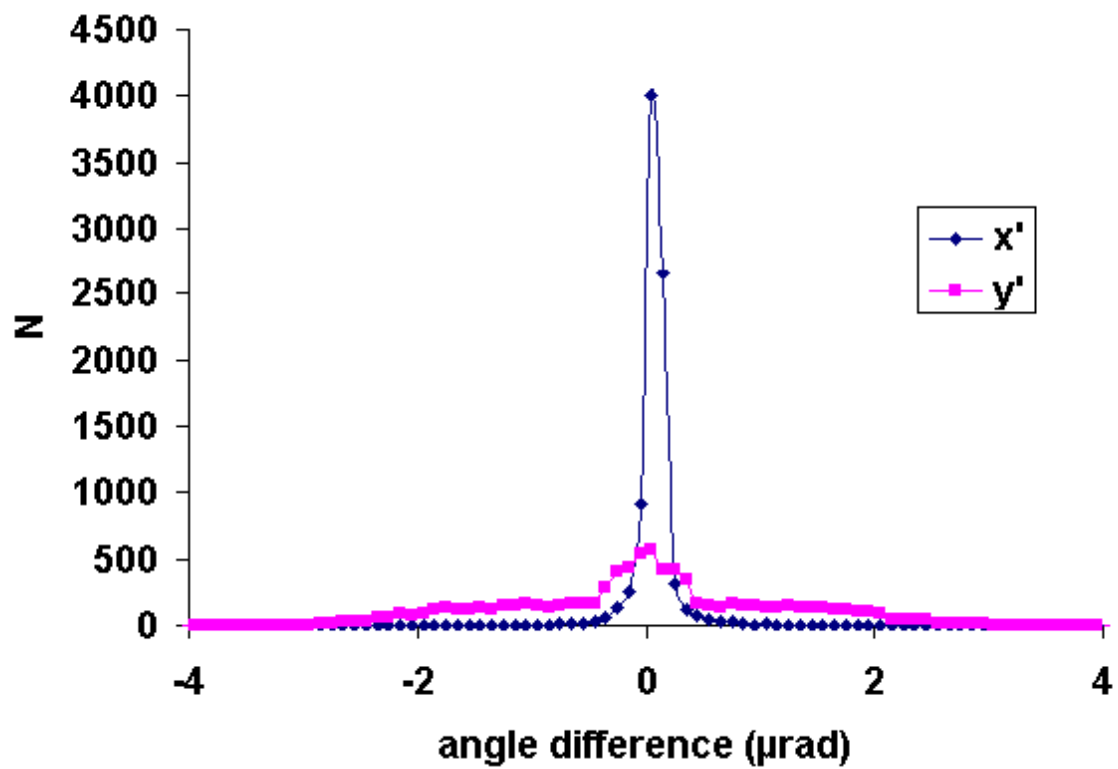

**Figure 1** The difference in proton angles, $x'$ and $y'$, before and after a Si(110) crystal.

Oscillations in the channeling plane on the atomic coherent potential are a greater problem. Fig. 1 shows a distribution of the difference in proton angle in $x'$ and $y'$ planes before and after a channeling in crystal, $(x'_{OUT} - x'_{IN})$ and $(y'_{OUT} - y'_{IN} - 0.1 \text{ mrad})$, as obtained in simulations for a Si(110) bent crystal channeling in $y'$ plane. The accuracy in $x'$ transmission in crystal is very good indeed, ~0.1 μrad rms. The width of $(y'_{OUT} - y'_{IN} - 0.1 \text{ mrad})$ distribution is much greater due to oscillations in the potential of Si (110) planes.

## 4. Precise transmission in both planes

To solve the problem of accuracy in the other plane, i.e. the plane of channeling, one solution is to use a channel with a lower critical angle, for instance Si(100) instead of (110) or (111). A more universal solution is to use a strongly bent crystal. The critical angle $\theta_C$ is gradually reduced to zero when the crystal curvature approaches a critical value. The strong focusing of a strongly bent crystal suppresses channeling oscillations to any low level needed in the application.

Fig. 2 shows the difference in proton angle in $x'$ and $y'$ planes before and after channeling in a crystal, ($x'_{OUT} - x'_{IN} - 0.1$ mrad) and ($y'_{OUT} - y'_{IN}$), as obtained in simulations for a 2 mm Si(100) crystal bent 0.1 mrad. The protons were channeled in $x'$ plane. The rms value of angle smearing found in simulations is 0.2 µrad both for $x'$ and $y'$.

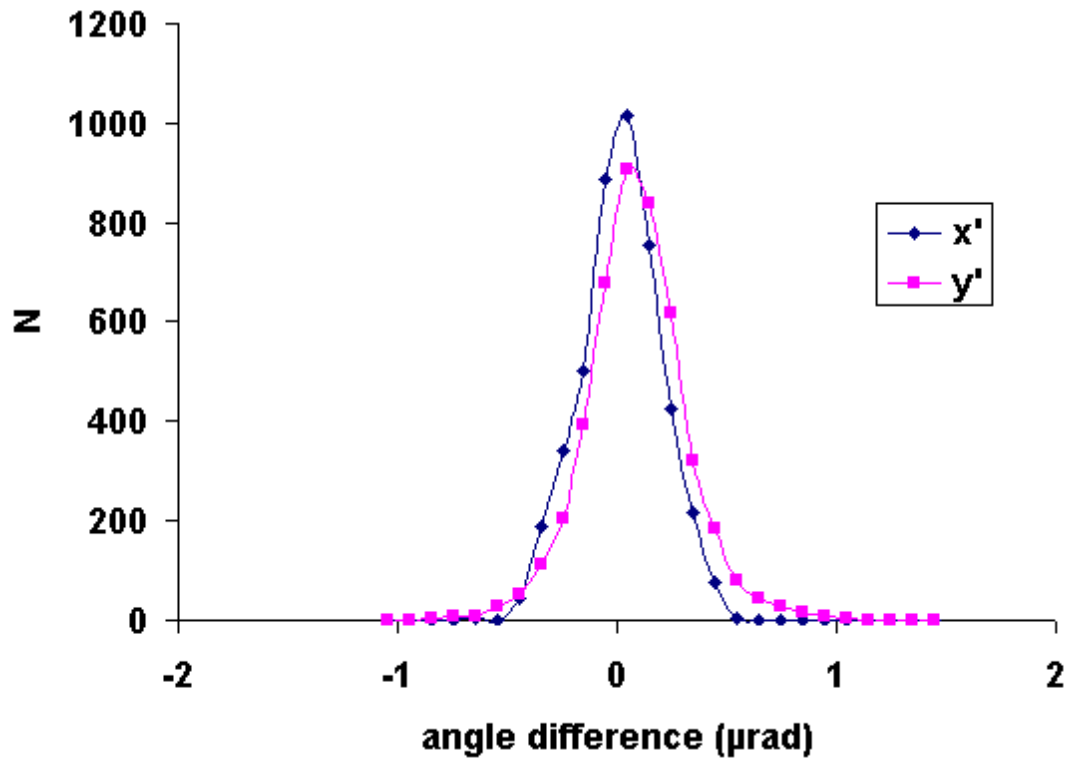

**Figure 2** The same as in Fig. 1 but for a strongly bent Si (100) crystal.

This accuracy should be compared to the angular resolution of the detectors downstream of the crystal. Measuring proton coordinates with ~10 µm resolution [13] over a base of ~8 m as allowed by a drift space would give an angular accuracy of ~1.4×10µm/8m=1.8 µrad. Addition (quadratic) of crystal transmission accuracies doesn't change this resolution. That would be perfect for crystal. With a much better resolution on the detector side, down to 0.5 µrad, the overall resolution becomes ~0.55 μrad, i.e. just slightly disturbed. Crystal transmission in both planes, *x'* and *y'*, is still almost perfect.

Because of scattering on electronic gas, the protons loose energy in crystal. In simulations, the energy loss and its fluctuations in crystal are $\Delta E/E \approx 10^{-7}$–$10^{-6}$, i.e. much smaller than even the nominal energy spread in the LHC beam, $1.1\times10^{-4}$. Energy losses in bent crystals were studied in experiments at CERN SPS with protons of 450 GeV and Pb ions of 33 TeV where CATCH predictions were also validated [9]. The diffractively scattered protons would have energy spread on the order of 100 GeV, or $\Delta E/E \approx 1.5\%$, at the crystal entrance. In simulations with $\beta^*$=0.55-2 m optics, channeling efficiency was completely independent of energy even for $\Delta E/E \approx 10\%$. In high $\beta^*$ options, crystal efficiency was uniform within ≈0.7% for $\Delta E/E \approx 1.5\%$. One can say that a phase space distribution (*x, y, x', y', E*) can be perfectly preserved in crystal and no information is lost on transmission in crystal.

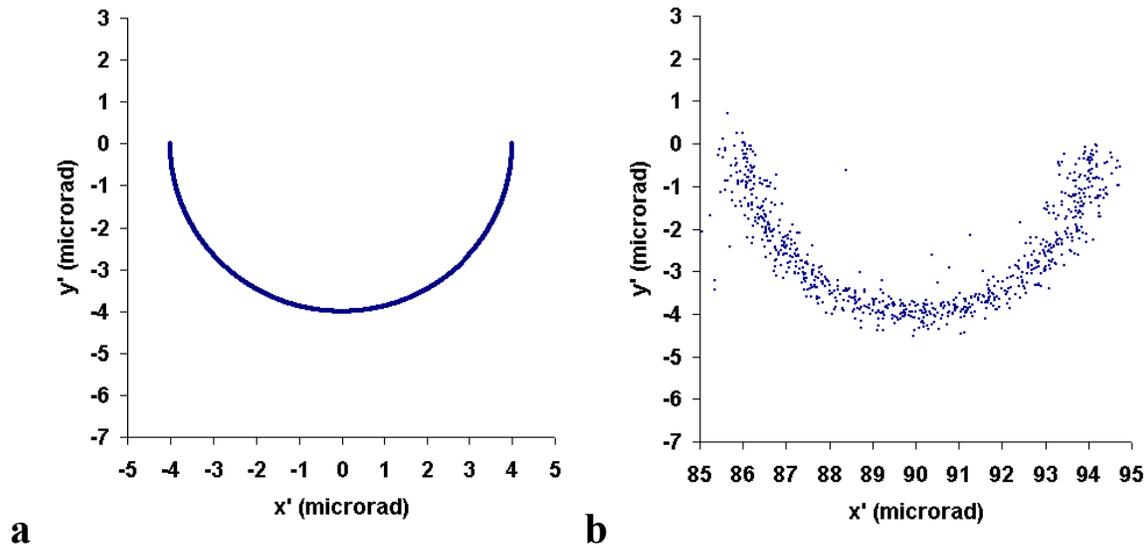

**Figure 3** An example of beam space (*x', y'*) at the entrance to the crystal (a) and at the exit (b).

Fig. 3 shows an example of a (*x', y'*) plot at the entrance to the crystal (a) and at the exit of it (b) where we tried to show how accurately a crystal can transmit a signature in angular space (semicircle chosen as a probe). The resolution of the image transmitted by a crystal is ~0.2 μrad in both planes. In terms of the critical channeling angle $\theta_C$, the obtained resolution is an order of magnitude finer than $\theta_C$ while the size of the trapped and channeled area can be some orders of magnitude greater than $\theta_C$.

In the applications there is no point to have a crystal transmission too perfect. It should match the other sources of inaccuracy like a multiple scattering in the detectors and vacuum chambers, etc. By tuning crystal parameters, in principle, one could very much improve in precision of the beam image downstream of the crystal but loose in brightness of the image, i.e. in statistics rate, as the efficiency of crystal transmission could be affected.

Finally, we suggest another idea for the channeling plane (a "microscope idea") that improves not only the crystal accuracy but even the detector resolution in that plane. Crystal can magnify beam image in one plane, e.g. transform the entrance values ($x, x', y, y'$) into exit values ($x, Nx' +\theta, y, y'$). The magnification factor $N$ can be as big as 2 or 10 or even 100, and serve the purpose to increase strongly the overall angular resolution in $x'$. In the above examples, the overall resolution was ~1 μrad defined by detector resolution. With magnification optics, the overall inaccuracy in $x'$ would be effectively reduced by factor $N$, bringing it below 0.1 μrad rms. Magnification is realized by making the shape of the crystal exit face different from the entry face. With a magnification factor of 10, e.g., the entry angular opening of 50 μrad would correspond to the exit opening of 500 μrad.

## 5. Conclusion

We have shown in simulations that crystal lattice can trap with 90% efficiency a beam with a ($x', y'$) distribution much broader than a critical angle $\theta_C$. To achieve that, one has to match the crystal focus length to the effective length between the particle source and the crystal in the accelerator lattice. Crystal adaptation to accelerator lattice improved channeling efficiency up to 300-fold. Crystal can transmit the trapped particles in channeled states with the phase space ($x, x', y, y', E$) distribution preserved with accuracy an order of magnitude finer than $\theta_C$. Several solutions were proposed and supported by simulations for achieving a fine resolution in crystal transmission.

This may give a beam instrument for collision products in colliders. Usually, accelerator beam instruments prepare particles for collision: by cooling them, bending, focusing, etc. Detectors sort out the results of collision. We change this a bit by introducing crystal optics between the collision point and detectors.

A crystal adapted to the LHC lattice can trap with 90% efficiency all protons emerging from the IP with divergence of 150 μrad or ~100$\theta_C$. The trapped protons can be channeled to detectors with precision down to 0.1 μrad rms. This makes feasible a crystal application for the measurement of diffractive scattering in CMS and ATLAS at the LHC. While we showed the physical capabilities of crystal channeling, its actual application in the LHC environment has to take into account many technical considerations to fit into existing infrastructure of accelerator and detectors.

Crystal channeling of LHC forward protons can improve proton acceptance in momentum loss $\xi$ and four-momentum transfer $t$ both in TOTEM and FP420 and allow to reach the smallest possible value of the scattering angle [9]. Now the sensitive detector area starts at ~12-15σ from the LHC beam [10]. Crystal can be placed at ~6σ from the LHC beam as it is very small, ~cm Si, and does not provoke beam instability. Such a crystal can trap and deliver a very useful information on most forward high momentum "quasi-elastic" and elastic protons at LHC, unavailable otherwise.

There are practical benefits as well. Crystal would relax tough requirements on $\beta^*$ needed for TOTEM. Crystal may allow TOTEM to run at the early start of the LHC, possibly running in parallel to other experiments. Thanks to crystal, FP420 detectors could possibly reside out of the cold region. The detectors don't need to be edgeless. Crystal works best with low $\beta^*$, where FP420 is interested most. If detectors can be more distanced from the beam, background conditions may improve. For injection, the active areas of the detectors must be kept away from the beams and then moved back; instead, one can move a crystal. Crystal can be introduced to experiment on a later stage in an attempt to expand the horizons of the physics program.